\documentstyle[amssymb,preprint,prl,aps]{revtex}
\begin{document}
\title{DESCRIPTION OF THE PRIMARY RELAXATION IN SUPERCOOLED LIQUIDS THROUGH THE
\smallskip TIMESCALE STEEPNESS FUNCTION. }
\author{{\em \ }Valery B. Kokshenev$^{1}$, Pablo D. Borges$^{1}$, and Neil S.
Sullivan$^{2}$}
\address{$^1$Departamento de Fisica, Universidade Federal de Minas Gerais, \\
C.P. 702, 30123-970, Belo Horizonte, Minas Gerais, Brazil\\
$^{2}$Physics Department, University of Florida, PO Box 117300, Gainesville,%
\\
FL 32611-8400}
\date{\today}
\maketitle

\begin{abstract}
\leftskip 54.8pt \rightskip 54.8pt

The primary relaxation in glass forming supercooled liquids (SCLs) above the
glass transformation temperature $T_{g}$ is discussed in terms of the
first-order (steepness) and the second-order (curvature) temperature
derivatives of the observed primary relaxation timescale. 
% are deduced from the experiments. 
We report new insights into the problem of the domain of the
Vogel-Fulcher-Tamman (VFT) equation, raised by Stickel et al. (J. Chem.
Phys. {\bf 102}, 6251 (1995), {\it ibid}. {\bf 104}, 2043 (1996)) and
discussed by Richert and Angell ({\it ibid.} {\bf 108}, 9016 (1998)). A new
ergodic-cluster Gaussian statistical approach to the problem is given based
on Onsager's thermodynamic principle. The primary relaxation is described by
the VFT equation below the crossover temperature $T_{c}$ (known from mode
coupling theory (MCT)), and above $T_{c}$ by an extended (VFTE) equation
obtained after accounting for cluster-size fluctuations. The timescale is
parametrized by a finite number of observable parameters such as the
steepness function ($m_{T}$) at $T_{g}$ % (i.e., by the fragility $m_{g}$)
, the MCT slowing-down exponent $\gamma _{c}$, and the VFT and VFTE strength
indices $D_{g}$ and $D_{c}$. The latter are defined at $T_{g}$ and $T_{c}$,
respectively, for the strongly and moderately SCL states, which show
absolute thermodynamic instability at the same VFT temperature $T_{0}$,
associated with the Kauzmann temperature. For both states the limiting
cluster-size characteristics are derived from experiment. A
thermodynamic-dynamic correspondence is established between the dynamic VFT
equation and the thermodynamic Adam and Gibbs model. As a by-product,
self-consistent equations for the excess heat capacity, entropy, and for the
glass-transition characteristic temperatures $T_{0}$, $T_{g}$, and $T_{c}$
are deduced. The problem of the ''irregular'' SCLs, which are not consistent
with the standard VFT equation, such as salol, ortho-terphenyl, and
bis-methoxy-phenyl-cyclohexane, is also discussed.\newline
PAC Nos: 00.00.+g, 00.00.Cn, 00.00.Rh 00.00.+g, 00.00.Cn, 00.00.Rh
\end{abstract}

\pacs{00.00.+g, 00.00.Cn, 00.00.Rh 00.00.+g, 00.00.Cn, 00.00.Rh}

\section{INTRODUCTION}

The understanding of noncrystalline liquids and the formation of glasses by
a dynamical slowing-down, and a rapid increase in viscosity at the glass
transition temperatures $T_{g\text{ }}$(near $2/3$ of the melting point $%
T_{m}$) is currently seen as a major intellectual challenge in condensed
matter physics \cite{ANM00}. There is therefore intense interest in studies,
experimental and theoretical, including simulations, of the temperature
dependence of the transport properties of supercooled liquids (SCLs). In
particular, data on the primary relaxation times $\tau _{T}^{(\exp )}$ that
are determined from {\em dynamic } experiments that study viscoelastic,
dielectric, conductivity, light scattering and mechanical relaxation data of
glass formers \cite{BNA93}, offer one of the main keys to the problem\cite
{ANM00,A91,GG92,Bin99}.

Unlike the slow secondary relaxation process which has a constant {\em %
activation energy} $\approx 24k_{B}T_{g}$ (established in Ref.%
\onlinecite{KTB98} for the Arrhenius relaxation, where $k_{B}$ is
Boltzmann's factor), the thermally equilibrated SCL states, observed within
the temperature domain $T_{g}\leq T<T_{m}$, exhibit a pseudo-Arrhenius form,
i.e., $\tau _{T}=\tau _{\infty }\exp (E_{T}/k_{B}T)$, and are discussed
traditionally in terms of the SCL activation energy $E_{T}$. The most widely
employed Vogel-Fulcher-Tammann (VFT) timescale fitting phenomenological form
with the activation energy $E_{T}=B/(T-T_{0})$ is introduced by the fictive
critical temperature $T_{cr}$ ($=T_{0}$, VFT temperature). $T_{0}$ lies
below the {\em thermodynamic} glass transition temperature $T_{g}$
established by {\em thermodynamic experiments }such as scanning calorimetric
or volume-expansion data. Analysis of the temperature-derivative behavior of
the $\alpha $-relaxation timescale $\tau _{T}$ given by Stickel et al. in
Refs. \cite{S95,S96,S97}, as improved and extended by Richert and Angell's
analysis of the thermodynamic data in Ref.\onlinecite{RA98}, determined the
real temperature observation domain, or {\em observation window}, of the VFT
form. Overall $\tau _{T}$ fitting forms, distinct from the VFT, were
proposed within the whole SCL observation window by Kivelson et al. in
Ref.11 or by Schulz in Ref.\onlinecite{Sch98} that, respectively, admits the
''avoided critical points''\cite{KKZ95} ($T_{cr}>T_{m}$), or rules out the
non-zero critical temperatures ($T_{cr}=0$).

Within the context of first-and-second-order phase-transition approaches to
the problem of the liquid-to-glass transition (see Refs.\cite
{AG65,CG79,WS87,SSH91,NRP91,Oda95,Bak98,Cha99,FB99,K99}) the dramatic
increase of the relaxation times $\tau _{T}^{(\exp )}$ observed in SCLs upon
cooling, is due to the growth of underlying correlation-length scales. The
lengthscale peculiarities, even when being experimentally established, do
not often lead to a clear physical understanding. In the case of the VFT
form the physical meaning of the experimentally determined VFT temperature $%
T_{0}$, is justified by its closeness to the {\em thermodynamic} Kauzmann
temperature $T_{K}$. At this critical temperature ($T_{K}=T_{cr}$) the
excess liquid-over-solid SCL entropy, or configurational entropy, vanishes.
This property is thought to be associated with the absolute thermodynamic
instability of the thermally equilibrated states of the SCLs with respect to
liquid-to-solid rearranging processes that extend from the ergodic region ($%
T_{g}\leq T<T_{m}$) to lower temperatures. The theoretically motivated
ergodic instability (caused by divergence of the zero moment of the relevant
relaxation-time distribution function\cite{Oda95}) establishes the
thermodynamic-dynamic (TD-D) correspondence equation, $%
T_{K}^{(TD)}=T_{0}^{(D)}\equiv T_{0}$. This equation was experimentally
justified by Angell\cite{A91} within a certain range of experimental scatter
which often lies\cite{A97} inside the error of experimentally measured
characteristic temperatures.

Thermodynamic models for SCLs are commonly tested by the $\tau _{T}^{(\exp
)} $ fitting analysis. The VFT form has proved to be ineffective for
establishing simultaneously all three adjustable parameters ($\tau _{\infty
}^{(VFT)}$, $B$ and $T_{0}$) within the overall fitting analysis of the
observed SCL viscosity (see discussion in Ref.\onlinecite{KTZ96}). An
improved analysis for the tested forms was proposed in Refs.\cite
{S95,S96,S97} on the basis of fitting experimental data $\tau _{T}^{(\exp )}$%
and additionally their first-order ($d\log \tau _{T}^{(\exp )}/dT$ ) and
second-order ($d^{2}\log \tau _{T}^{(\exp )}/dT^{2}$) derivatives. These
temperature-derivative analyses permit one to exclude the unknown fitting
parameters, step-by-step, by making the three-parameter fitting procedure
self-consistent. As a result, the validity of the VFT form, when tested
within the whole ergodic relaxation range, was questioned\cite{S97} for all
SCLs investigated above the so-called $\alpha $-$\beta $-relaxation
bifurcation temperature $T_{B}$. Notably, this crossover temperature was
often found close to other known {\em crossover temperatures }$T_{x}$ and $%
T_{c}$ introduced, respectively, as a scaling temperature by R\"{o}ssler and
co-workers\cite{NRM96} and as the dynamic critical temperature of the mode
coupling theory (MCT). Within these studies one can admit that to a good
approximation, $T_{B}=T_{x}=T_{c}$ (see e.g. Table II in Ref.%
\onlinecite{RA98}).

Physically, the MCT crossover temperature $T_{c}$ signals structural
transformations, or liquid-solid-like restructuring in SCLs, from{\em \ }%
moderately supercooled states of ''weakly coupled fluids'' ($T_{c}\leq
T<T_{m}$) to strongly supercooled states of ''strongly coupled fluids'' ($%
T_{g}\leq T<T_{c}$) as discussed in Ref.\onlinecite{GG92}. The analyses
given in Refs.\cite{S95,S96,S97,RA98} in terms of the first order
derivatives (related to the timescale steepness) and the second-order
derivatives (related to the timescale curvature) of the temperature
dependence, clearly distinguish the two regions, similar to the case of the
MCT theory [4], below and above the dynamical crossover temperature $T_{B}$,
associated here with $T_{c}$. The data for moderately SCL (or {\em weak-SCL}%
) states in all glass forming liquids that have been studied are in a good
quantitative agreement with the modified VFT form given as the {\em high-T}
VFT (HT-VFT) version in the old canonical VFT form but with new adjustable
parameters, including the HT-VFT characteristic temperature $T_{0}^{(HT)}$($%
\approx T_{g} $, see Table II in Ref.\onlinecite{S96}). This fundamental
finding is exemplified by the case of the simple molecular liquid propylene
carbonate (PC) in {Fig.1}. As seen from the analysis given in Fig.1, the
validity of the canonical VFT form is limited by the temperature domain $%
T_{g}\leq T<T_{c}$, corresponding to the strongly SCL states, characterized
by relaxation times $10^{-7}\lesssim \tau _{T}^{(\exp )}/s\leq 10^{2}$, or
by viscosities $10^{4}\lesssim \eta _{T}^{(\exp )}/P\lesssim 10^{13}$.

Another interesting description of the SL states is also provided by
analysis of the thermodynamic (heat capacity) data that was elucidated by
Richert and Angell\cite{RA98} on the basis of the {\em configurational
entropy} $S_{c}$ data. This data is schematically plotted in {Fig.2}.

Careful complex dynamic-thermodynamic analyses enabled one to establish a
physical meaning for the standard VFT fitting form, that was justified in
terms of the thermodynamic Adam-Gibbs model (AGM) introduced in Ref.%
\onlinecite
{AG65}. The VFT form is therefore treated as the {\em VFT equation} that
provides a good description of the observed dynamic and thermodynamic
relaxation of the {\em strong-SCL} states (see AG-line in Fig.2). Within the
context of percolation theory\cite{Isi92}, $T_{0}$ may be associated with
the critical temperature where the correlation length of the strong-SCL
states diverges in accordance with the thermodynamic instability at the
Kauzmann temperature T$_K$. A microscopic version of this kind of
instability was given by Bender and Shlesinger Ref.\onlinecite{BS98} by
invoking a defect-aggregation mechanism that was recently justified \cite
{BFS01} for PC by the pressure-dependent experiments.

In reality, strong-SCL ergodic states do not survive below $T_{g}$ (see
solid squares in Fig.2). As shown in Ref.\onlinecite{K01}, this is due to
the ergodic-to-nonergodic crossover from the short-range strong liquid
correlations, characteristic of solids, to somewhat long-range weak liquid
correlations, characteristic of gases. Conversely, no physical justification
was given for the HT-VFT form introduced in Refs.\cite{S95,S96,S97,RA98}.
The main aim of the current study is to provide a physical description of
the weak-SCL states on the basis of both the extended thermodynamic AGM and
the dynamic VFT equations.

The old idea that anomalous temporal (non-Debye) and anomalous temperature
(non-Arrhenius) behaviors related to the SCL relaxation, may be attributed
to rearranging regions\cite{FB99}, or domains\cite{Cha99}, or droplets\cite
{Bak98} or clusters\cite{K01} of different size and structure, is still
widely explored and remains fruitful. There is today considerable direct
evidence for the existence of these kinds of regions. They have been
observed on a microscopic scale by multidimensional NMR, photobleaching,
excess light scattering and dielectric hole burning, as reviewed in Ref.%
\onlinecite
{ANM00}, and by X-ray scattering data, discussed in Ref.\onlinecite{FB99}.
The percolation theory, improved by mesoscopic-domain energy-size scaling,
as shown by Chamberlin in Ref.\onlinecite{Cha96}, provided a description in
detail of the observed relaxation function. An alternative
percolation-theory approach, combined with the relaxation-time cluster-size
scaling law, proposed in Ref.\onlinecite{K98} shed light on some universal
features of the primary relaxation near $T_{g}$ in the structural and
orientational glasses\cite{KS01J}. Additional cluster-approach studies of
the dynamical spectra with accounting of the MCT scaling forms, explained
the mechanisms\cite{KS01} for the primary relaxation in polymers and those
of the secondary relaxation in SCLs. Furthermore, exploration of Stauffer's
cluster-size scaling form, based on the percolation theory, made it possible%
\cite{K01} to justify the underlying ergodic-nonergodic transition that
emerges at the ergodic crossover temperature $T_{E}$ ($\le T_{g}$), where
the cluster-size Gaussian distribution of the strong-SCL states is
transformed upon cooling into the Poisson form (see also discussions in Refs.%
\cite{Oda95,CBS92}).

In the current study we extend the cluster approach to the problem of
relaxation of ergodic states in SCLs. We will demonstrate that the
structural heterogenous fluctuations, being small in the case of the
strong-SCL state, are responsible for moderately SCL behavior as revealed in
Refs.\cite{S95,S96,S97,RA98} by deviations from the standard VFT fitting
form. The paper is organized as follows. In Sec. II we discuss
phenomenological descriptions of the observed primary relaxation timescale
in terms of the known fitting forms and macroscopic parameters commonly used
by experimentalists. Instead of the frequency observation windows, given by $%
1/\tau _{T}^{(\exp )}$, which are distinct for different dynamic
experiments, the timescale is introduced through the temperature domains for
the SCL excitations established with the help of the characteristic
temperatures. In Sec. III we give the dynamic and thermodynamic description
of SCLs based on the statistical approach to the rearranging regions
introduced by Adam and Gibbs. By accounting for their size fluctuations, we
develop extended versions for the AGM and the VFT equations. The qualitative
and quantitative analyses of the proposed physically justified fitting forms
are given in Sec. IV. The conclusions are summarized in Sec. V.

\section{MACROSCOPIC PARAMETRIZATION OR THE PRIMARY RELAXATION TIMESCALE}

\subsection{Strongly Supercooled Liquids}

It was experimentally proved by Richert and Angell in Ref.\onlinecite{RA98}
that the widely employed AGM equation

\begin{equation}
\log _{10}\tau _{T}^{(AG)}=A+\frac{C}{TS_{c}(T)}  \label{tau1-AG}
\end{equation}

\noindent and the phenomenological VFT form given in the two equivalent
forms, namely 
\begin{eqnarray}
\log _{10}\tau _{T}^{(VFT)} &=&\log _{10}\tau _{\infty }^{(VFT)}+\frac{B}{%
T-T_{0}}\text{ ,}  \label{tauB-VFT} \\
\tau _{T}^{(VFT)} &=&\tau _{\infty }^{(VFT)}\exp \frac{DT_{0}}{T-T_{0}}
\label{tauD-VFT}
\end{eqnarray}
are self-consistent when subject to the conditions $A=\log \tau _{\infty
}^{(VFT)}$ and $T_{0}=T_{K}$ . These equations apparently work well in
describing the strong-SCL state relaxation in terms of the MCT within the
temperature domain $T_{g}\leq T<T_{c}$. This domain was studied\cite{RA98}
simultaneously on the basis of dynamic and thermodynamic experimental data,
more precisely, through the primary relaxation timescale $\tau _{T}^{(\exp
)} $, deduced from the dielectric loss spectra, and the configurational
entropy:

\begin{equation}
S_{c}(T)\equiv \Delta S_{T}=\int_{T_{K}}^{T}\Delta C_{T}T^{-1}dT  \label{Sc1}
\end{equation}
with the excess liquid-over-solid {\em isobaric} specific heat capacity $%
\Delta C_{T}=C_{p}^{(Liq)}-C_{p}^{(Sol)}$. Accounting for the fact that at
high enough temperatures\cite{NM67,Pri80} $\Delta C_{T}^{(\exp )}\propto
T^{-1}$, the configurational entropy (\ref{Sc1}) was given\cite{RA98,AS76}
in explicit form, namely 
\begin{equation}
S_{c}(T)=\Delta S_{\infty }\left( 1-\frac{T_{K}}{T}\right) \text{ with }%
\Delta S_{\infty }=\frac{C}{B}  \label{Sc2}
\end{equation}
that establishes useful relations between the observed model parameters. It
has been demonstrated\cite{RA98} that unlike the VFT form, restricted by the
strong-SCL states, Eq.(\ref{Sc2}) does a good job within the whole range of
the thermally equilibrated states (see solid line in Fig.2).

The theoretical foundation of the experimental findings by Richert and
Angell are not limited by the aforesaid TD-D correspondence equation\cite
{RA98}, i.e., $T_{K}^{(TD)}=T_{0}^{(D)}\equiv T_{0}$. One can see that the
existence of the strongly SCL states, based on the dynamic VFT equation and
the thermodynamic AGM within observation windows, evidently requires in
addition $T_{g}^{(TD)}=T_{g}^{(D)}\equiv T_{g}$. Here the characteristic
temperature $T_{g}^{(TD)}$ is introduced as a thermodynamic parameter
derived from the maximum of $C_{p}^{(Liq)}(T_{g}^{(TD)})$, observed by the
scanning calorimetry, and the corresponding laboratory glass-point $%
T_{g}^{(D)}$ that is deduced from the dynamic experiments and defined\cite
{BNA93} by the equation $\tau _{T}^{(\exp )}=10^{2}s$ given at $%
T=T_{g}^{(D)} $. Unlike the case of the first TD-D equation, there is no
systematic study to verify the precision of the second equation. Except for
the case of SCLs (see e.g. Fig.2 in Ref.\onlinecite{A95}) there is even no
accepted international convention on the definition of the thermodynamic
value $T_{g}^{(TD)}$ for the various glass forming materials.

Furthermore, Eqs.(\ref{tau1-AG}, \ref{tauB-VFT}) cannot be extrapolated to
high temperatures beyond their temperature domain. Hence, the justification
given in Ref.\onlinecite{RA98} for the VFT and the ADM corresponding
macroscopic equations, i.e., $B=C/\Delta S_{\infty }$ and $A=\log \tau
_{\infty }^{(VFT)} $, should be reconsidered. The same equations might be
introduced by application of the {\em ergodic hypothesis} employed for SCLs
in Ref.\onlinecite
{K01}. This means that the strongly SCL states, that emerge above the
crossover-ergodic-temperature $T_{E}$ (see Eq.(7) in Ref.\onlinecite{K01}),
can be introduced through certain observables estimated by thermodynamic and
dynamic averages. Physically, this corresponds to the fact that the
thermally equilibrated states are equivalent to ergodic states observed in
the dynamical experiments. If this is true, this requires new equations for
the ergodic time scale and length scale given, respectively, by 
\begin{equation}
\tau _{\infty }^{(VFT)}=\tau _{\infty }^{(AG)}\equiv \tau _{a}\text{ and }%
z_{\infty }^{(VFT)}=z_{\infty }^{(AG)}\equiv z_{a}.  \label{tau-and-z}
\end{equation}
In other words, Eq.(\ref{tau-and-z}) introduces the minimum magnitudes for
temporal and size characteristics necessary employed for the mesoscopic
description of the SCL states.

\subsection{Moderately Supercooled Liquids}

In contrast to the second-order thermodynamic transition approach, MCT
treats the liquid-to-glass structural transformation as a dynamic process
driven by liquid density fluctuations. In the idealized version of MCT [4]
the ergodic-to-nonergodic transition occurs at a unique crossover
temperature $T_{c}(=T_{cr})$ at which neutron spin scattering measurements
signal the appearance of the glassy-like correlations given by the so-called
non-ergodic parameter, which is analogues to the Edwards-Anderson spin-glass
order parameter in the case of weak-SCL states. Moreover, the MCT\ predicts%
\cite{GG92} the algebraic-type scaling for the primary relaxation timescale,
namely 
\begin{equation}
\tau _{T}^{(MCT)}=\frac{\tau _{\infty }^{^{(MCT)}}}{(T-T_{c})^{\gamma _{c}}}%
\text{ for }T_{c}<T_{y}\leq T<T_{A}<T_{m}  \label{tau-MCT}
\end{equation}
with the slowing-down exponent $\gamma _{c}$. This observable is rather
sensitive to the location of $T_{c}$ as well to the temperature region of
the fitting experimental curve\cite{GG92,S95,S96,KTZ96,Sou90}. The
temperature domain of Eq.(\ref{tau-MCT}) is much narrower than that for the
case of the HT-VFT form given\cite{S97} by $T_{B}<T<T_{A}$, where $T_{A}$ is
the Arrhenius crossover temperature. A temperature $T_{y}$ is introduced in
Eq.(\ref{tau-MCT}) to designate the lowest temperature point where the
algebraic MCT form fits experimental data (see Fig.1). Unlike the first
estimates reported for the slowing-down exponents that exceeded magnitudes
of $5.5$ in SCLs (see Table1 in Ref.\onlinecite{Sou90}), the values
established by the MCT Eq.(\ref{tau-MCT}) in glass-forming materials are in
the range, $2<\gamma _{c}<3$ for liquids Eq.(\ref{tau-MCT}), and $3\leq
\gamma _{c}<4$ for polymers. For macroscopic observable parameters of the
timescale in glass forming molecular liquids and polymers see, respectively,
cases A and B in {Table 1}.

\subsection{Fragility and Characteristic Temperatures}

\smallskip One of the salient features of the primary relaxation is the
so-called {\em time-temperature} superposition principle\cite{GG92}. Its
physical justification has not yet been established but it is employed
effectively to scale the dielectric loss spectra\cite{GG92}. Meanwhile,
analyses of a number of the $\alpha $-timescale characteristics near $T_{g}$
for several experimental studies of glass forming materials\cite{BNA93},
have revealed a number of underlying {\em temporal-spatial} correlations in
explicit form. The analysis was given\cite{BNA93} in terms of the
phenomenological relation, i.e., $m_{g}=(250-320)\beta _{g}$ expressed
through the {\em fragility}\cite{BNA93} $m_{g}$ that is related to the
steepness index\cite{RA98}, and the Kohlrausch-Williams-Watts (KWW)
stretched exponent\cite{ANM00} $\beta _{g}$.

Incorporation\cite{K98} of the growth-correlation scaling law, employed in
the order-disorder kinetic transition theory\cite{Bra94}, into the
relaxation function of the SCLs led to a description\cite{KS01J} of the
aforesaid temporal-spatial correlations for the dynamical exponents in a
form $\beta _{g}=(1+$ $m_{g}/m_{0}^{*})^{-1}$ where $m_{0}^{*}$ is a
universal parameter (because it is material-independent) within a certain
glass family. In addition, the equation for the growth-correlation dynamic
exponent, i.e., $z_{g}=m_{g}/3m_{0}^{*}$ , was established in Ref.\,[33](see
also proposal by Ngai in Ref.\onlinecite{Nga94}).

Within MCT, the characteristic temperatures, as well as the dynamical
exponents, are treated as certain equilibrium thermodynamic parameters,
which, as demonstrated in Ref.\onlinecite{K99}, are not independent. This
statement is exemplified by the widely employed equation\cite{fragile}

\begin{equation}
\frac{T_{g}}{T_{0}}=\frac{m_{g}}{m_{g}-m_{g}^{*}}\text{, with }%
m_{g}^{*}=\log _{10}\frac{\tau _{g}^{*}}{\tau _{a}}\text{ ,}  \label{Tg/T0}
\end{equation}
where $m_{g}^{*}=16\pm 2$ is the universal parameter known as the {\em lower
fragility limit }\cite{BNA93}. As follows from Eq.(\ref{Tg/T0}), the
uncertainty in the $m_{g}^{*}$, caused\cite{K99,precise} by deviation of the
inclination of the straight lines in {Fig.\,3}, is due to the uncertainty of
the timescale unit parameter $\tau _{\infty }^{(VFT)}=\tau _{a}=10^{-14\pm
1}s$, given by Eqs.(\ref{tau-and-z}), and to that of the glass-transition
characteristic time $\tau _{g}=\tau _{g}^{*}=10^{2\pm 1}s$. Eq.(\ref{Tg/T0})
was first introduced in Refs.\cite{BNA93,BA92} on the basis of the {\em VFT
equation} (\ref{tauD-VFT}), namely 
\begin{equation}
\tau _{T}^{(VFT)}=\tau _{\infty }^{(VFT)}\exp \frac{D_{g}}{\varepsilon _{T}}%
\text{ with }\varepsilon _{T}=\frac{T}{T_{0}}-1\text{ for }T_{g}\leq T<T_{c},
\label{tau-VFT-equation}
\end{equation}
where the {\em timescale auxillary function} $\varepsilon _{T}$ is
introduced.

Let us define the {\em timescale} {\em steepness function} by

\begin{equation}
m_{T}=-\frac{d\log _{10}\tau _{T}}{d\ln T}\text{ }  \label{steep-func}
\end{equation}
with 
\begin{equation}
m_{T}^{(VFT)}=\frac{D_{g}}{\ln 10}\frac{1}{\varepsilon _{T}}(1+\frac{1}{%
\varepsilon _{T}})  \label{mT-VFT}
\end{equation}
for the particular case of Eq.(\ref{tau-VFT-equation}). Eq.(\ref{mT-VFT}),
estimated at $T=T_{g}$, is known\cite{BA92} as the fragility $m_{g}$ that
was proposed by Angell for the strong-to-fragile classification of glass
forming materials\cite{define}. We see that unlike the case of $\tau
_{T}^{(VFT)}$, the function $m_{T}^{(VFT)}$ has only two free parameters, $%
T_{0}$ and $D_{g} $. Moreover, the latter can be excluded from the
two-equation system given by $\tau _{g}^{(VFT)}$and $m_{g}^{(VFT)}$, using $%
D_{g}=m_{g}^{*}\varepsilon _{g}\ln 10$, that eventually results in the known
Eq.(\ref{Tg/T0}).

As shown above, Eq.(\ref{Tg/T0}) is a formal consequence of the
self-consistency between the VFT equation $\tau _{g}^{(VFT)}$ and its first
derivative given by the observed fragility $m_{g}^{(VFT)}$. Physically, the
existence of Eq.(\ref{Tg/T0}) is due to the presence of the strong-SCL
states at the lowest temperature $T_{g}$ parametrized by these equations.
One can therefore introduce a similar relation (see Eq.8 in Ref.%
\onlinecite{K99}), namely

\begin{equation}
\frac{T_{c}}{T_{0}}=\frac{m_{g}+m_{c}^{*}}{m_{g}-m_{c}^{*}}\text{ with }%
m_{c}^{*}=\log _{10}\frac{\tau _{c}^{*}}{\tau _{a}}\text{ }  \label{Tc/T0}
\end{equation}
associated with the weak-SCL states at the lowest temperature $T_{c}$. The
validity of Eq.(\ref{Tc/T0}) was established numerically with the universal
parameter $m_{c}^{*}=7\pm 1$ (see Fig. 2 in Ref.\onlinecite{K99}) that
corresponds to the adopted value $\tau _{c}^{*}=10^{7\pm 1}s$. Eq.(\ref
{Tc/T0}) was deduced\cite{K99} by application of the Taylor series for the
steepness function (\ref{steep-func}) near $T_{g}$ and $T_{c}$ along with
the following assumptions: (i) in contrast to the case of the timescale $%
\tau _{T}$ the steepness function $m_{T}$ is analytical and smooth beyond $%
T_{0}$; and (ii) similar to the strong-SCL state, described by $m_{T}^{(VFT)}
$ with parameters $D_{g}$ and $T_{0}$, the weak-SCL states can be described
within the scope of the HT-VFT form with help of $m_{T}^{(HT)}$, using $%
D_{c}^{(HT)}=m_{c}^{*}\stackrel{\sim }{\varepsilon }_{c}\ln 10$, and $%
\stackrel{\sim }{\varepsilon }_{c}=T_{c}/T_{g}-1$, i.e. by the formal
substitution of $T_{0}$ and $T_{g}$ in the standard VFT equation by $T_{g}$
and $T_{c}$ in the HT-VFT form, respectively. It is remarkable that the idea
to introduce in Ref.\onlinecite{K99} the steepness function $m_{T}$ on the
basis of the HT-VFT analytical form was stimulated by the uncertainty of the
fitting analyses of $\tau _{T}^{(\exp )}$ discussed by Kivelson et al. in
Ref.\onlinecite{KTZ96}. Experimentally, the HT-VFT form was established by
the first-order derivative analyses given by Stickel et al. in Refs.\cite
{S95,S96,S97}.

The most striking consequence of the existence of the two distinct SCL
states described at $T_{g}$ (in the vicinity but below $T_{c}$),
respectively, by Eqs.(\ref{Tg/T0}) and (\ref{Tc/T0}) is an inference on the
self-consistency of the principal characteristic temperatures. Indeed,
excluding the fragility $m_{g}$ from Eqs.(\ref{Tg/T0}) and (\ref{Tc/T0}) one
obtains the following {\em characteristic-temperature equation }(CTE),
namely 
\begin{equation}
\frac{T_{c}}{T_{g}}=\frac{T_{g}\left( \frac{m_{g}^{*}}{m_{c}^{*}}+1\right)
-T_{0}}{T_{g}\left( \frac{m_{g}^{*}}{m_{c}^{*}}-1\right) +T_{0}},
\label{Kok}
\end{equation}
(that was analyzed analytically in Ref.\onlinecite{K99} under the model
simplification $m_{g}^{*}=2m_{c}^{*}$). One can see that beyond the scope of
this simplification, Eq.(\ref{Kok}) includes the particular case of the MCT 
{\em ideal} liquid-to-glass transition that occurs at a unique crossover
temperature given by the equation $T_{g}=T_{0}=T_{c}$.

We have shown that representation of the CTE in the form given in Eq.(\ref
{Kok}) is ensured by application of the VFT fitting\ forms to both the
strong and the weak SCL states. On the other hand, there exist a number of
fitting forms (see e.g. Refs.\cite{BS98,BFS01} and for a short review see
Ref.\onlinecite{ANM00}) and therefore the question arises as to whether the
fact of self-dependency of characteristic temperatures given through the CTE
is also limited by exploration of the VFT forms.

Using the Lindeman melting criterion for the mean-square displacements of
pre-transition solid-like clusters in SCLs, Novikov et al. in Ref.%
\onlinecite
{NRM96} deduced the CTE in the ''square-root'' form: $T_{g}^{(low)}=\sqrt{%
T_{0}T_{c}}$. Meanwhile it was shown\cite{K99} that $T_{g}^{(low)}$ and the
''cubic-root'' CTE, i.e., $T_{g}^{(up)}=$ $^{3}\sqrt{T_{0}T_{c}^{2}}$, are
respectively, the lower and the upper limits of the solutions of Eq.(\ref
{Kok}) considered under the model constraint $m_{g}^{*}=2m_{c}^{*}$.
Physically, this means that the experimental data on $T_{g}^{(\exp )}$ and
theoretical predictions for $T_{g}$, deduced from Eq.(\ref{Kok}) under
conditions $T_{0}=T_{0}^{(\exp )}$ and $T_{c}=T_{c}^{(\exp )}$, are expected
to satisfy the equation $T_{g}=T_{g}^{(\exp )}$ within the experimental
error of $\pm 5\%$ (see the last column in Table 1 in Ref.\onlinecite{K99})
and, in addition, to satisfy the inequalities $%
T_{g}^{(low)}<T_{g}<T_{g}^{(up)}$.

Odagaki proposed a controlled-diffusion approach to the SCL glass-transition
problem in Ref.\onlinecite{Oda95}. A glass transformation process was
characterized by the critical VFT temperature $T_{0}$ ($=T_{K}$) and the
crossover temperature $T_{c}$ ($=T_{x}$) related to the peculiarities of the
pseudo-divergent $\alpha $-timescale, and introduced\cite{Oda95} through the
true divergence of the first moment of their distribution function. As a
consequence the CTE in the form $T_{c}\Delta S_{c}=2T_{g}\Delta S_{g}$ was
deduced. Taking into account that the excess entropy $\Delta S_{T}$, given
in Eq.(\ref{Sc1}), in real SCLs is well approximated by Eq.(\ref{Sc2}), we
represent here the Odagaki CTE in the following equivalent form, namely 
\begin{equation}
\frac{T_{g}}{T_{c}}=\frac{1}{2}\left( 1+\frac{T_{0}}{T_{c}}\right) \text{.}
\label{Oda}
\end{equation}
Again, one can see that the aforementioned MCT ideal transition is also
satisfied by Eq.(\ref{Oda}). A comparative analysis of the distinct CTE form
is given in {Fig.4} for glass forming liquids (case A) and polymers (case
B). As seen from Fig.4 the Odagaki CTE prediction (\ref{Oda}) is very close
to that given by Eq.(\ref{Kok}) with a value for $m_{g}^{*}/m_{c}^{*}$ that
lies within the experimentally established uncertainties, i.e., $%
m_{g}^{*}=16\pm 2 $ and $m_{c}^{*}=7\pm 1$.

\section{DYNAMICS AND THERMODYNAMICS OF SUPERCOOLED LIQUID STATES}

\subsection{Cluster Statistics}

Solid-like clusters were treated in Refs.\cite{K01,KS01J,KS01} as
short-range {\em random-size} correlated regions of SCLs that permits one,
at a {\em mesoscopic} level of consideration, to employ the percolation\cite
{K01,K98} or controlled-diffusion\cite{KS01J,KS01} models. Within this
context, clusters are equivalent to solid-like domains\cite{KKZ95,Cha96}, or
droplets\cite{Bak98} or CRRs (cooperatively rearranging regions) as
introduced in Ref.\,\cite{AG65}. A more detailed specification of clusters
depends on ways of incorporation of short-range correlations in the
description of their evolution.

We treat clusters as conglomerates of $z$ molecules\cite{cluster} that are
characterized at a given temperature $T$ and pressure $p$ by the solid-state
chemical potential $\mu _{T}^{(sol)}(p)$. These {\em finite-size} solid-like
clusters are in thermal and mechanical contact with the remaining $N-z$
molecules, which are in the normal liquid (NL) state and characterized by a
chemical potential $\mu _{T}^{(liq)}(p)$. The typical cluster, or {\em %
characteristic cluster}, is in dynamic equilibrium with SCL, and is
introduced by the mean-size {\em configurational average}, i.e., $%
<z>_{C}=z_{T}(p)$ through the cluster-size distribution function $P_{T}(p,z)$%
. For the thermally equilibrated SCL states we admit the ergodicity
hypothesis\cite{K01}, i.e. the configurational averaged values are equal to
those estimated at thermodynamic equilibrium for temperatures $T>T_{g}\geq
T_{E}$, above the ergodic-nonergodic transition temperature T$_E$ [29].

The {\em isobaric} process, as a model mechanism of $z$-cluster formation,
can be described by variation of the Gibbs (thermodynamic) potential $\delta
G^{(sol)}(T,p,z)=-S_{T}^{(sol)}\delta T+\mu _{T}^{(sol)}\delta z$.
Solidification, or cluster growth, if assumed at the thermal equilibrium
temperature $T_{eq}$, is defined by minimization of the whole system
potential, that gives rise to the requirements imposed on the {\em chemical
potentials}, i.e., $\mu _{eq}^{(sol)}=\mu _{eq}^{(liq)}$, and on the {\em %
entropies}, with $S_{eq}^{(sol)}<S_{eq}^{(liq)}$, given at $T=T_{eq}$ (see
e.g. Fig.\,15 and discussion in Ref.\onlinecite{L-L}). Stabilization of the
liquid-solid boundary upon decrease of temperature below the equilibrium
temperature is also ensured by minimization of the whole Gibbs potential and
therefore by positivity of the excess chemical potential, i.e., $\Delta \mu
_{T}=\mu _{T}^{(sol)}-\mu _{T}^{(liq)}>0$ and the {\em mesoscopic-scale}
configurational entropy, i.e.,{\em \ }$\Delta
S_{T}(z)=S_{T}^{(liq)}(N-z)-S_{T}^{(sol)}(z)>0$ at $T<T_{eq}$ (hereafter we
omit the pressure $p$ as an irrelevant parameter). Note that the {\em %
macroscopic} configuration entropy introduced by Eq.(\ref{Sc1}) is defined
here as $\Delta S_{T}=<\Delta S_{T}(z)>_{C}$. The latter is also expected to
be valid far from thermodynamic equilibrium.

Within the {\em fluctuation }mechanism for the formation of heterogeneous
clusters, the probability for the formation of a solid cluster of size $z$,
is formally given by the probability $P_{T}(z)dz$ that the thermodynamic
variable $z$ lies between $z$ and $z+dz$. This probability is driven by the
temperature-dependent whole-system entropy\cite{onsager}, namely 
\begin{equation}
P_{T}(z)=\exp \frac{S_{T}(z)}{k_{B}}\approx \sqrt{\frac{2}{\pi \Delta
z_{T}^{2}}}\exp \left[ -\frac{(z-z_{T})^{2}}{2\Delta z_{T}^{2}}\right] .
\label{Pz}
\end{equation}
Here the characteristic cluster of size $z_{T}$ is in thermodynamic
equilibrium with the thermal bath, and therefore the entropy of the whole
system $S_{T}(z)=S_{T}^{(sol)}+S_{T}^{(liq)}+S_{T}^{(mix)}$ (which includes
the solid-liquid mixing term $S_{T}^{(mix)}$) has a maximum at $z=z_{T}$.
The {\em cluster-size fluctuations }$\Delta z_{T}^{2}\equiv
<(z-z_{T})^{2}>_{C}$ are introduced formally by the relation $\Delta
z_{T}^{2}=-(\partial ^{2}S_{T}/k_{B}\partial ^{2}z)_{z=z_{T}}^{-1}$, and Eq.(%
\ref{Pz}) is expected to work near equilibrium, i.e., for $T\approx T_{eq}$.
It is notable that the real SCLs, studied in dynamic experiments, are not
close to thermodynamic equilibrium, but the Gaussian size-distribution
function $P_{T}(z)$ of Eq.\thinspace (\ref{Pz}) was experimentally tested
and justified by Chamberlin et al. in Refs.\cite{Cha96,CBS92} for a number
of SCLs.

Let us treat the formation of a cluster of size $z$ at fixed pressure $p$
and temperature $T$ as an {\em isobaric-isothermal fluctuation} {\em process}%
. This means that solid-like clusters are not in thermodynamic equilibrium
with the fluid, i.e., $T<T_{eq}$, ($T$ refers to the solid-state subsystem
and $T_{eq}$ to the thermal bath) but they appear and disappear from a fluid
volume of size $z$ with a CRR{\em \ frequency} $\tau _{T}^{-1}(z)$, where $%
\tau _{T}(z)$ is the lifetime of a $z$-cluster. The probability of formation
of this kind of cluster is given by $\tau _{T}^{-1}(z)$ and is determined
from Onsager's principle, i.e. by the aforesaid mesoscopic size-scale
entropy $\Delta S_{T}(z)$, namely

\begin{equation}
\tau _{T}^{-1}(z)=\tau_{\infty}^{-1}\exp \left( \frac{\Delta S_{T}(z)}{k_{B}}%
\right) =\tau_{\infty}^{-1}\exp \left[ -\frac{W_{\min }(z)}{k_{B}T_{eq}}%
\right] \text{.}  \label{tau-1}
\end{equation}
Here $W_{\min }=\Delta G_{T}(z)$ corresponds to the {\em minimum work} (see
e.g. Eqs.(20.8) and (114.1) of Ref.\onlinecite{L-L}) required for
solidification of $z$ molecules driven by cluster-size fluctuations. Within
the isobaric-isothermal mechanism, proposed by Adam and Gibbs\cite{AG65},
one has $\Delta G_{T}(z)=(\mu _{T}^{(sol)}-\mu _{_{eq}}^{(liq)})z$. As a
result the exact result of Eq.(\ref{tau-1}) can be extended to
non-equilibrium states with $T\approx T_{eq}$ through the relation

\begin{equation}
\tau _{T}(z)=\tau _{a}\exp \left( \frac{\Delta \mu _{T}\text{ }z}{k_{B}T}%
\right)  \label{tauTz}
\end{equation}
which introduces the cluster {\em mesoscopic-scale relaxation time}. This in
turn is related to the observed {\em macroscopic timescale}, namely

\begin{equation}
\tau _{T}\equiv <\tau _{T}(z)>_{C}=\int_{0}^{\infty }\tau _{T}(z)P_{T}(z)dz%
\text{.}  \label{tauT}
\end{equation}
Within the proposed cluster statistical treatment, the corresponding{\em \
macroscopic lengthscale} is given by 
\begin{equation}
z_{T}\equiv <z>_{C}=\int_{0}^{\infty }zP_{T}(z)dz\text{.}  \label{z-T-defin}
\end{equation}
On one hand, these two scales, experimentally justified above the glass
transition temperature, are ensured by the thermodynamic-dynamic
correspondence principle. On the other hand, the cluster-formation
mechanism, specified by $P_{T}(z)$ through the observables given in Eqs.(\ref
{tauT}) and (\ref{z-T-defin}), should satisfy the constraints that are
naturally required by the macroscopic and mesoscopic scales. In other words,
the observables should obey the conditions $\tau _{T}>\tau _{T}^{(\min
)}=\tau _{a}$ and $z_{T}>z_{T}^{(\min )}= z_{a}$ where the limiting
mesoscopic-scale magnitudes $\tau _{a}$ and $z_{a}$ are established in Eq.(%
\ref{tau-and-z}).

\subsection{Adam-Gibbs Model}

The Adam-Gibbs molecular-kinetic theory\cite{AG65} was based on Eq.(\ref
{tau-1}) given for the cooperative transition probability for rearranging
regions of $z$ molecules with the minimum work $W_{\min }^{(AG)}=\Delta \mu
^{(AG)}z$ that was estimated for the {\em isobaric-isothermal }ensemble of
equivalent thermodynamic subsystems, where $\Delta \mu ^{(AG)}$ is a
solid-over-liquid excess chemical potential, approximated by a constant. The
average transition probability $1/\tau _{T}^{(AG)}$ follows from the AGM
macroscopic timescale, namely

\begin{equation}
\tau _{T}^{(AG)}=\tau _{\infty }^{(AG)}\exp \left( \frac{\Delta \mu ^{(AG)}%
\text{ }z_{T}^{(AG)}}{k_{B}T}\right) .  \label{tau2-AG}
\end{equation}
The smallest CRR of a ''critical size'' $z_{T}^{(AG)}$, that still permits a
transition, corresponds to the lowest {\em critical} configurational entropy
for one CCR, $s_{c}^{*}=k_{B}\ln 2$, introduced as the first estimate in Ref.%
\cite{AG65}. The configurational entropy $S_{c}(T)$ of the whole system
given in Eqs. (\label{Sc1,Sc2}\ref{Sc1}) and (\ref{Sc2}) for one mole of the
SCL ($N=N_{A}$), consists of $N_{A}/z_{T}^{(AG)}$ of self-similar smallest
CRRs and is therefore estimated here as 
\begin{equation}
S_{c}(T)=s_{c}^{*}\frac{N_{A}}{z_{T}^{(AG)}}.  \label{Sc-AG}
\end{equation}
One can see that Eq.(\ref{Sc-AG}) is equivalent to Eq.(20) in Ref.%
\onlinecite{AG65}. Eqs.(\ref{tau2-AG}) and (\ref{Sc-AG}) deduced above
result in the known Eq.(\ref{tau1-AG}) with the AGM parameters, namely 
\begin{equation}
A=\log _{10}\tau _{\infty }^{(AG)}\text{ and }C=\frac{\Delta \mu
^{(AG)}N_{A}s_{c}^{*}}{k_{B}}.  \label{A,C-AG}
\end{equation}
Note that the AGM was motivated to validate the phenomenological
Williams-Landel-Ferry fitting form that in a way is equivalent to the VFT
form.

Adopting the aforesaid TD-D\ correspondence established between the AGM and
the VFT form given by Eq.(\ref{tauB-VFT}), and with the help of Eqs.(\ref
{Sc2}) and (\ref{tau2-AG}), we introduce the primary relaxation lengthscale
through the relations 
\begin{equation}
z_{T}^{(AG)}=\frac{z_{\infty }^{(AG)}}{1-T_{K}/T}\text{ with }z_{\infty
}^{(AG)}=z_{\infty }^{(VFT)}=z_{a}=\frac{N_{A}s_{c}^{*}}{\Delta S_{\infty }}=%
\frac{k_{B}B}{\Delta \mu ^{(AG)}\ln 10}  \label{CRR-T}
\end{equation}
where the minimum cluster-size magnitude $z_{a}$ is anticipated by Eq.(\ref
{tau-and-z}).

The CRR size-temperature behavior given by $z_{T}^{(\exp )}$ was
experimentally studied for SCLs\cite{model} and for orientational-glass
plastic crystals in Refs.\cite{TYM95,YIT99}. These studies as well as the
findings of Ref.\onlinecite{RA98}, both based on the AGM, clearly indicate
that Eq.(\ref{CRR-T}), and also Eq.(\ref{Sc2}), do good a job within the
whole SCL ergodic temperature range, i.e., at $T\geq T_{g}$ (see e.g. Fig. 4
in Ref.\onlinecite{TYM95} and Fig. 3 in Ref.\onlinecite{RA98}). This permits
one to present the experimental result for the excess isobaric heat capacity 
$\Delta C_{T}^{(\exp )}\propto T^{-1}$, used above to evaluate the
configurational entropy in Eq.(\ref{Sc2}), in the following equivalent form: 
\begin{equation}
\Delta C_{T}^{(\exp )}=\Delta C_{g}^{(\exp )}\frac{T_{g}}{T}\text{ for }%
T\geq T_{g}.  \label{deltaC-T}
\end{equation}
With the help of Eq.(\ref{Sc1}) one arrives at a new useful relation for the
observed model parameter $\Delta S_{\infty }$ (see Table I in Ref.%
\onlinecite{RA98}) introduced in Eq. (\ref{Sc2}), namely 
\begin{equation}
\Delta S_{\infty }=\Delta C_{g}^{(\exp )}\frac{m_{g}}{m_{g}-m_{g}^{*}}
\label{delta-Sin}
\end{equation}
that though needs further experimental examination.

Finally, one can see that the CRRs can be introduced by the SCL cluster
statistics on the basis of Eq. (\ref{tauT}) with the help of $%
P_{T}^{(AG)}(z)=\delta (z-z_{T}^{(AG)})$. This means that the CRRs are {\em %
homogeneous clusters} of characteristic size $z_{T}=$ $z_{T}^{(AG)}$ given
in Eq.(4) with negligible heterogeneous and/or thermal size fluctuations,
i.e., $\Delta z_{T}=\Delta z_{T}^{(AG)}=0$.

\subsection{Extension of the Vogel-Fulcher-Tamman Equation}

Within the cluster statistics approach the primary relaxation timescale is
introduced by Eqs.(\ref{Pz}),(\ref{tauTz}) and (\ref{tauT}). A
straightforward calculation using the standard integral\cite{GR} 
\begin{equation}
\int_{0}^{\infty }\exp \left( -\frac{x^{2}}{4b}-ax\right) dx=\sqrt{\pi b}%
\left[ 1-\text{erf}\left( {a\sqrt{b}}\right) \right] \exp ba^{2}  \label{GR}
\end{equation}
yields 
\begin{equation}
\tau _{T}=\tau _{a}\left[ 1+\text{erf}\left( \frac{1+\Delta E_{T}/k_{B}T}{%
\sqrt{2\Delta E_{T}/E_{T}}}\right) \right] \exp \left[ \frac{E_{T}}{k_{B}T}%
\left( 1+\frac{\Delta E_{T}}{2k_{B}T}\right) \right]  \label{tau-E}
\end{equation}
given in the pseudo-Arrhenius form with the cluster-energy characteristics,
namely 
\begin{equation}
E_{T}=\Delta \mu _{T}z_{T}\text{, }\Delta E_{T}=\frac{\Delta \mu _{T}\Delta
z_{T}^{2}}{z_{T}}\text{, }z_{T}=\frac{z_{a}}{1-T_{0}/T}\text{.}  \label{ET}
\end{equation}
The standard error function that emerges in Eq.(\ref{tau-E}) is bounded by
one, and although it modifies the pre-exponential factor $\tau _{a}$, it
does not substantially modify the exponential temperature dependence of $%
\tau _{T} $. Furthermore, the error-function term will be absorbed by $\tau
_{a}$.

We see that the VFT form of Eq.(\ref{tau-E}), when extended by cluster-size
fluctuations, has the form of a high-$T$ thermodynamic perturbation series.
The first and second terms in the exponent describe the rigid solid clusters
and their fluctuation corrections, respectively. The first term matches the
rigid-cluster relaxation time given by the VFT equation and the AGM solution
given by Eq.(\ref{tau-VFT-equation}) and Eq.(\ref{tau2-AG}), respectively.
The TD-D\ correspondence is now extended at a mesoscopic level that permits
one to establish the physical meaning of the macroscopical phenomenological
parameters for the strong-SCL states, namely 
\begin{equation}
B^{(SL)}=\frac{\Delta \mu _{g}z_{a}}{k_{B}\ln 10}\text{ and }C^{(SL)}=\frac{%
\Delta \mu _{g}N_{A}s_{c}^{*}}{k_{B}\ln 10}  \label{B,C-SL}
\end{equation}
under the additional condition $\Delta \mu _{g}=\Delta \mu ^{(AG)}$ along
with Eq.(\ref{tau-and-z}).

Unlike the case of the thermodynamic AGM, which gives a good description\cite
{RA98} of all the SCL states (see Fig.2) within the ergodic temperature
domain, $T_{g}\leq T\leq T_{m}$ (that means that $C=C^{(SL)}=C^{(ML)}$ in
Eq.(\ref{tau1-AG})), the VFT form fails\cite{S95,S96,S97} to describe weak
SCL states if one uses the same critical temperature $T_{0}$. In order to
incorporate the weak SCL states into the $\alpha $-timescale parametrized by
the observed MCT parameters ($\tau _{\infty }^{^{(MCT)}}$, $T_{c}$, and $%
\gamma _{c}$) let us represent the exponential form (\ref{tau-E}) in the
asymptotically equivalent algebraic form given by Eq.(\ref{tau-MCT}). Using $%
\exp \left[ c\left( 1+x\right) \right] \equiv \exp \left[ (1+x)^{c}\right] $%
, one has the following high-$T$ series with respect to the small
fluctuation parameter $\Delta E_{T}/k_{B}T$ , namely

\begin{eqnarray}
\tau _{T}^{(VFTE)} &=&\tau _{a}\left[ \exp \left( 1+\frac{\Delta E_{T}}{%
2k_{B}T}\right) \right] ^{\frac{E_{T}}{k_{B}T}}= \\
\tau _{a}\left[ e\left( 1+\frac{\Delta E_{T}}{2k_{B}T}+\frac{1}{2}\left( 
\frac{\Delta E_{T}}{2k_{B}T}\right) ^{2}+...\right) \right] ^{\frac{E_{T}}{%
k_{B}T}}\text{ } &\propto &\frac{\tau _{\infty }^{^{(MCT)}}}{\left( 1-\frac{%
T_{c}}{T}\right) ^{\gamma _{c}}}\text{ for }T>T_{c}  \label{tau-VFTE}
\end{eqnarray}
denoted the VFT {\em extended} (VFTE) form. Accounting for only the two
leading terms, one establishes the following relations, i.e., $\gamma
_{c}=E_{c}/k_{B}T_{c}$ and $T_{c}=\Delta E_{c}/2k_{B}$ with $\tau _{\infty
}^{(MCT)}=e^{\gamma _{c}}\tau _{a}$, that can also be represented with the
help of Eq.(\ref{ET}) in terms of the observed MCT parameters, namely 
\begin{equation}
\gamma _{c}=\frac{\Delta \mu _{c}z_{\infty }^{(VFTE)}}{k_B T_{0}\varepsilon
_{c}}=\frac{2}{\xi _{c}^{2}}\text{ with }\varepsilon _{c}=\frac{T_{c}}{T_{0}}%
-1\text{ }  \label{gama-c}
\end{equation}
and

\begin{equation}
T_{c}=T_{0}+\frac{\Delta \mu _{c}z_{\infty }^{(VFTE)}}{k_B \gamma _{c}}.\;
\label{Tc}
\end{equation}
Here a new model parameter $\xi _{c}=\sqrt{\Delta E_{T}/E_{T}}=\Delta
z_{T}/z_{T}$ is introduced\cite{K01}, with the aim of establishing a scale
for the cluster-size fluctuations above $T_{c}$. Accounting for the fact
that the size fluctuations are limited, at least to $\xi _{c}< 1$, with help
of Eq.(\ref{gama-c}) one arrives at a physical restriction $\gamma _{c}>2$
that is in accord with experimental data. As a result, the $\alpha $%
-relaxation scale is now parametrized for the strong SCL states, namely 
\begin{equation}
\tau _{T}^{(SL)}=\tau _{a}\exp \left( \frac{\Delta \mu _{g}z_{a}^{(SL)}}{%
k_{B}T_{0}\varepsilon _{T}}\right) \text{ with }\varepsilon _{T}=\frac{T}{%
T_{0}}-1\text{, for }T_{g}\leq T<T_{c}\text{,}  \label{tau-SL}
\end{equation}
with the observable combinations of cluster parameters $\Delta \mu
_{g}z_{a}^{(SL)}=D_{g}k_{B}T_{0}$ (see Eq.(\ref{tau-VFT-equation})) and $%
z_{a}^{(SL)}=z_{\infty }^{(VFT)}$ (see also Eq.(\ref{tau-and-z})). For the
case of the weakly cooled states in moderately SCLs one has 
\begin{equation}
\tau _{T}^{(ML)}=\tau _{a}\exp \left[ \frac{\Delta \mu _{c}z_{a}^{(ML)}}{%
k_{B}T_{0}\varepsilon _{T}}\left( 1+\frac{1}{\gamma _{c}}\frac{\Delta \mu
_{c}z_{a}^{(ML)}}{k_{B}T_{0}\varepsilon _{T}}\right) \right] \text{ for }%
T_{c}\leq T\leq T_{m}  \label{tau-ML}
\end{equation}
with $\Delta \mu _{c}z_{a}^{(ML)}=\gamma _{c}k_{B}(T_{c}-T_{0})$ that
follows from Eq.(\ref{Tc}) when $z_{\infty }^{(VFTE)}=z_{a}^{(ML)}$. In the
following sections we review experimental evidence for the proposed
description of the primary relaxation through the timescale steepness and
curvature.

\section{STEEPNESS FUNCTION ANALYSIS}

\subsection{Qualitative Analysis}

We discuss two scenarios of relaxation of the SCLs observed in thermodynamic
and dynamic experiments and shown in Fig.\,2. The thermodynamic version is
given by the AGM that describes the configurational entropy data within the
whole SCL ergodic range (shown by solid squares in Fig.2). Within the
context of percolation treatment, the solid clusters are apparently the same
for the strongly and moderately SCL states, i.e., $%
z_{a}^{(SL)}=z_{a}^{(ML)}(=z_{\infty }^{(AG)}$) and thermal fluctuations are
negligible ($\Delta z_{T}^{(AG)}=0$). The AGM excess chemical potential is
approximated by a constant given at $T_{g}$, i.e., $\Delta \mu
^{(AG)}=\Delta \mu _{g}$.

To describe the primary relaxation timescale observed in the dynamic
experiments, we start from the trial estimate given by the thermodynamic
approximation. This is formally introduced by SCL-system clusters with $%
z_{a}^{(SCL)}=z_{\infty }^{(AG)}=z_{a}$, but with $\Delta z_{T}^{(SCL)}\neq
0 $, that corresponds to an extension of the AGM by cluster-size
fluctuations, which strictly speaking are not observed in real thermodynamic
experiments. Meanwhile the trial dynamic time-scale steepness function
follows from Eqs.(\ref{steep-func}) and (\ref{tau-ML}), namely

\begin{eqnarray}
m_{T}^{(SCL)} &=&\frac{D^{(SCL)}}{\ln 10}\frac{1}{\varepsilon _{T}}\left( 1+%
\frac{1}{\varepsilon _{T}}\right) \left( 1+\frac{2}{\gamma _{c}}\frac{%
D^{(SCL)}}{\varepsilon _{T}}\right) \text{ with}  \nonumber \\
\text{ }D^{(SCL)} &=&\frac{\Delta \mu _{g}z_{a}}{k_{B}T_{0}}\text{ for }%
T_{g}\leq T<T_{m}\text{,}  \label{m-SCL}
\end{eqnarray}
where the auxiliary timescale function $\varepsilon _{T}$ is given in Eq.(%
\ref{tau-VFT-equation}). In order to employ the timescale parametrization in
terms of the observed fragility $m_{g}$, we apply Eq.(\ref{m-SCL}) at $%
T=T_{g}$. Using the relation $z_{a}=k_{B}T_{0}\gamma _{c}\varepsilon
_{c}/\Delta \mu _{c}$, one excludes the unknown parameter $z_{a}$. By
solving Eq.(\ref{m-SCL}), with respect to the MCT parameter $\gamma _{c}$,
one obtains the trial estimate given by 
\begin{equation}
\gamma _{c}=\frac{m_{g}^{*}\ln 10}{\frac{\Delta \mu _{g}}{\Delta \mu _{c}}%
\frac{\varepsilon _{c}}{\varepsilon _{g}}\left( 1+2\frac{\Delta \mu _{g}}{%
\Delta \mu _{c}}\frac{\varepsilon _{c}}{\varepsilon _{g}}\right) }\text{ }
\label{gama-cTA}
\end{equation}
that was analyzed in detail in Ref.\onlinecite{K01}. Here we consider Eq.(%
\ref{gama-cTA}) as the quadratic equation with respect to the unknown factor 
$\Delta \mu _{g}/\Delta \mu _{c}$. This enables one to calculate the
rebuilding energy per molecule near the crossover temperature, namely 
\begin{equation}
\frac{\Delta \mu _{g}}{\Delta \mu _{c}}=\frac{\varepsilon _{g}}{4\varepsilon
_{c}}\left( \sqrt{1+8\frac{m_{g}^{*}}{\gamma _{c}}\ln 10}-1\right) \text{.}
\label{mu-g/mu-c}
\end{equation}
The physical solutions of the quadratic equation (\ref{gama-cTA}) must
satisfy the requirement of thermodynamic stability discussed above and given
by the condition $\Delta \mu _{g}\geq $ $\Delta \mu _{c}$. Numerical
solutions of Eq.(\ref{mu-g/mu-c}) are shown in {Table 2} for experimental
data known for the characteristic temperatures and the slowing-down exponent 
$\gamma _{c}^{(\exp )}$ given in Table 1. Predictions for the unknown $%
\gamma _{c}$ are also made with the help of Eq.(\ref{gama-cTA}) through the
overall estimates $\Delta \mu _{g}^{(SCL)}/\Delta \mu _{c}^{(SCL)}=1.2$ and $%
\Delta \mu _{g}^{(pol)}/\Delta \mu _{c}^{(pol)}=1$ adopted, respectively,
for liquids and polymers. We see that no rebuilding chemical potential
effects are found for the case of glass forming polymers. This can be
understood as follows. In contrast to the case of molecular liquids, where
application of the {\em ergodic} cluster description by the Gaussian
distribution given in Eq.(\ref{Pz}) is justified both physically and
experimentally, the monomeric-segment polymers require accounting of their
fractal structure even for their ergodic states\cite{K01}.

The dynamic relaxation scenario given by the VFT {\em equation} is realized
through rigid clusters as well as by the VFT equation{\em \ }extended by the
cluster-size fluctuations (VFTE equation) and applied, respectively, to the
strong and weak SCL states. The relevant steepness function follows
immediately from the corresponding relaxation times. Hence with help of Eq.(%
\ref{tau-SL}), one has

\begin{equation}
m_{T}^{(SL)}=\frac{D_{g}}{\ln 10}\frac{1}{\varepsilon _{T}}(1+\frac{1}{%
\varepsilon _{T}})\text{ with }D_{g}=\frac{\Delta \mu _{g}z_{a}^{(SL)}}{%
k_{B}T_{0}}  \label{mT-SL}
\end{equation}
for the case of the strongly SCLs. In the case of the moderately SCLs the
steepness function follows from Eq.(\ref{tau-ML}), namely 
\begin{equation}
m_{T}^{(ML)}=\frac{D_{c}}{\ln 10}\frac{1}{\varepsilon _{T}}\left( 1+\frac{1}{%
\varepsilon _{T}}\right) \left( 1+\frac{2}{\gamma _{c}}\frac{D_{c}}{%
\varepsilon _{T}}\right) \text{with }D_{c}=\frac{\Delta \mu _{c}z_{a}^{(ML)}%
}{k_{B}T_{0}}.  \label{mT-ML}
\end{equation}

Qualitatively, the first-order derivative analyses of experimental data by
Stickel et al., given in Refs.\cite{S95,S96} points to the existence of a
small {\em kink of the steepness function} near the crossover temperature $%
T_{c}$ (see Fig.1)\cite{trial}. Formally, this means that the limit of the
timescale steepness function from below ($T\rightarrow T_{c}-0$ and $%
m_{T}^{(SL)}\rightarrow m_{c}^{(SL)}$) differs with that from above ($%
T\rightarrow T_{c}+0$ and $m_{T}^{(ML)}\rightarrow m_{c}^{(ML)}$ ), and
therefore one should expect that $m_{c}^{(SL)}>m_{c}^{(ML)}$. Application of
these limits to Eqs.(\ref{mT-SL}) and (\ref{mT-ML}) using the relation $%
D_{c}=\gamma _{c}\varepsilon _{c}$, that follows from Eq.(\ref{Tc}), results
in $D_{g}>3D_{c}$. These qualitative findings extended by the condition of
the thermodynamic cluster stability can be summarized as 
\begin{equation}
\frac{\Delta \mu _{g}}{\Delta \mu _{c}}\frac{z_{a}^{(SL)}}{z_{a}^{(ML)}}>3%
\text{ with }\frac{\Delta \mu _{g}}{\Delta \mu _{c}}>1\text{.}  \label{ineqs}
\end{equation}
The dynamic conditions given in Eq.(\ref{ineqs}) reflect the observed\cite
{S95,S96} dynamical crossover from moderately to strong SCLs near $T_{c}$
that are associated with the dynamic reconstruction of the cluster
lengthscale. To provide deeper insight into the problem we will reproduce
below the fitting derivative analyses by Stickel et al. within the VFT and
VFTE equations.

\subsection{Numerical Analysis}

\subsubsection{First-Order Derivative Timescale Steepness}

The first-order derivative analysis of the temperature behavior of the
timescale was given in Refs.\cite{S95,S96} with the help of a first-order
derivative function, namely 
\begin{equation}
F_{1T}=\left( -\frac{d\log _{10}\tau _{T}}{d1/T}\right) ^{-1/2}=\frac{1}{%
\sqrt{Tm_{T}}}.  \label{F1-T}
\end{equation}
This function was estimated from $\tau _{T}^{(exp)}$ and fitted by the VFT
form, i.e. by $F_{1T}^{(VFT)}=\left( 1-T_{0}/T\right) /\sqrt{B}$ , and by
other known fitting forms (see e.g. Fig.1). We reproduce this analysis in {%
Figs.\,5} and {6} in terms of the steepness function deduced from Eq.(\ref
{F1-T}), i.e., $m_{T}^{(\exp )}=1/T(F_{1T}^{(\exp )})^{2}$, with $%
F_{1T}^{(\exp )}$ found by Stickel et al. by fitting with $m_{T}^{(SL)}$ and 
$m_{T}^{(ML)}$. The latter are given, respectively, in Eq.(\ref{mT-SL}) and
Eq.(\ref{mT-ML}), with the {\em strength indices} $D_{g}$ and $D_{c}$
treated as adjustable parameters within the corresponding temperature
regions, $T_{g}\leq T<T_{c}$ and $T_{c}\leq T<T_{m}$, respectively. The
results of fitting analyses are accumulated in {Table 3}. Unlike the case of
Figs.\,5, where the VFT temperature $T_{0}$ is treated as the same for the
both states, the SCLs analyzed in Fig.\,6 are in conflict with the canonical
VFT form for the strongly SCL states, but this is in accord with analysis
given in Refs.\cite{S95,S96,RA98}. In this case the critical temperature $%
T_{0}$ is not established and is therefore treated as an additional
adjustable parameter for description of the moderately SCL states through
the VFTE equation. Thereby, we follow the idea that the Kauzmann temperature 
$T_{0}$ might be established beyond the VFT equation.

Our quantitative analysis of the strong-to-moderate-state crossover is in
accord with the qualitative predictions given by Eq.(\ref{ineqs}). As seen
from{\normalsize \ }Table 3, the typical change in the ratio of strength
indices $D_{g}$/$D_{c}$ lies between 4 and 6. Exceptions should be given for
the ''irregular'' case of salol where the Kauzmann temperature established
by the VFTE equation ($T_{0}=200K$) is rather different from those
established by other methods and given in Table 1.

Adopting the trial estimate for the moderate-to-strong SCL-state crossover
for the excess chemical potential, given in Table 2, one arrives at the
length-scale ratio estimate $z_{a}^{(SL)}/z_{a}^{(ML)}=3.7-4.7$. We see that
the observed crossover, associated with the cluster reconstruction, is not
caused solely by decrease of cluster-size fluctuations and/or by growth of
their chemical potentials. This crossover is accompanied by the lengthscale
extension, i.e., $z_{a}^{(SL)}>z_{a}^{(ML)} $.

\subsubsection{Second-Order Derivative Timescale Curvature}

The second-order timescale derivative-fitting analysis introduced by Stickel
et al. can be treated as experimental observation of the critical
temperature $T_{0}$ through the VFT form. This analysis was done\cite
{S95,S96} with the help of the second-order derivative function $%
F_{2T}(\equiv \Theta (T))$, introduced in Eq.(21) in Ref.\onlinecite{S95} as 
\begin{equation}
F_{2T}=\frac{d\log _{10}\tau _{T}}{dT}\left( \frac{d^{2}\log _{10}\tau _{T}}{%
dT^{2}}\right) ^{-1}=T\left( 1-\frac{d\ln m_{T}}{d\ln T}\right) ^{-1}
\label{F2-T}
\end{equation}
that results in $F_{2T}^{(VFT)}=-(T-T_{0})/2$ for the case $\tau _{T}=\tau
_{T}^{(VFT)}$ given in Eqs.(\ref{tauB-VFT}) and (\ref{tauD-VFT}). The
experimental data for $\tau _{T}^{(\exp )}$ was used to estimate $%
F_{2T}^{(\exp )}$ by Eq.(\ref{F2-T}) and fitted to the temperature-linear
function $F_{2T}^{(VFT)}$. As a result, the two distinct critical
temperatures $T_{0}$ and $T_c$ were established for PC and 1-propanol
through the VFT and HT-VFT forms (see Figs.\,13 and 16 in Ref.%
\onlinecite{S96}, respectively). In the case of salol and PDE, only the
HT-VFT form was derived\cite{S96} in Figs.\,8 and 11. Remarkably, in all
cases the assumed kink in the function $F_{1T}^{(\exp )}$ near $T_{c}$ was
transformed into the evident jump of the function $F_{2T}^{(\exp )}$.

The {\em curvature} of the timescale $k_{T}$, closely related to the
second-order derivative function given in Eq.(\ref{F2-T}), can be introduced
by the following relations, namely 
\begin{equation}
k_{T}\equiv \frac{d\ln m_{T}}{d\ln T}=\frac{1}{\ln 10}\frac{d^{2}\ln \tau
_{T}}{(d\ln T)^{2}}=1-\frac{T}{F_{2T}},  \label{curva}
\end{equation}
with $k_{T}^{(VFT)}=(3T-T_{0})/(T-T_{0})$. In order to describe the weakly
cooled states in moderately SCLs through the curvature $k_{T}^{(ML)}$ we
substitute the relevant steepness function given in Eq.(\ref{F2-T}). This
yields

\begin{equation}
F_{2T}^{(ML)}=-\frac{T}{1+\left( \frac{\gamma _{c}+4D_{c}}{\gamma _{c}+4D_{c}%
}+\frac{1}{\varepsilon _{T}+1}\right) \left( 1+\frac{1}{\varepsilon _{T}}%
\right) }\text{ with }D_{c}=\frac{\Delta \mu _{c}z_{a}^{(ML)}}{k_{B}T_{0}}
\label{F2-VFTE}
\end{equation}
that leads to an expression for $k_{T}^{(ML)}$ with the help of Eq.(\ref
{curva}). The typical SCL timescale curvatures are exemplified by PC and
salol given in, {Fig.\,7} and {Fig.\,8}, respectively. We see that both
cases are distinct and that the moderately SCL states are consistent with
the cluster-size fluctuation description, based on the idea of the existence
of a unique thermodynamic-instability temperature $T_{K}=T_{0}$. They are
described by the VFTE timescale equation (\ref{tau-ML}) with the same model
parameters established in the steepness timescale analysis given in Fig.\,5A
and 6A, respectively, for PC\ and salol. Conversely, in the case of salol
and PDE, our attempts failed\cite{thes} to the describe the strongly SCL
states through the observed curvature with the help of various VFT-type
fitting forms reviewed in Ref.\onlinecite{ANM00}, including the proposal
given in Refs.\onlinecite{BS98,BFS01}. Accounting for the aforesaid TD-D
correspondence, we speculate that the failure of the standard VFT equation
in the salol-type SCLs (that should be extended by the case of BMPC (see
Fig.6C)), is due to the inapplicability of the cluster description given by
the thermodynamic model of Adam and Gibbs in Ref.\onlinecite{AG65}. Indeed,
as shown above, the main AGM constraint for the excess chemical potential,
i.e., $\Delta \mu _{T}=\Delta \mu _{c}$, works well above the SL-ML
crossover temperature, but below $T_{c}$ it needs to be reconsidered for the
salol-type SCLs.

\section{DISCUSSION\ AND\ CONCLUSIONS}

The investigation of the temperature behavior of the primary relaxation
timescale in SCLs plays a fundamental role in understanding the glass
transition problem in general, as widely discussed in various,
microscopically distinct, glass forming non-solid materials. In spite of the
fact that the nature of the structural glass transformations is essentially
dynamical, the process of solidification of rearranging regions, while
avoiding crystallization of a given SCL as a whole system, exhibits many
features of the common liquid-to-solid or disorder-to-order true
thermodynamic phase transitions.

The ''pure'' dynamic treatment of the glass-transition problem, given by the
simplified MCT version, and the ''pure'' thermodynamic treatment given by
the simplified AGM, are limited in the description of the primary relaxation
timescale temperature behavior. This limitation was revealed by the
temperature derivative analyses given by Stickel et al., as shown in Fig.1,
that clearly establish that the fitting forms for the known $\tau
_{T}^{(MCT)}$ and $\tau _{T}^{(AG)}$ (given, respectively, in Eqs.(\ref
{tau-MCT}) and (\ref{tau-SL})), fit the data only for a restricted
temperature window.

The same kind of restriction applies to the analysis of the thermodynamic
models\cite{Sch98,KKZ95} mentioned in the Introduction, which offer a
similar overall $\tau _{T}$ description within the whole range of existence
of the ergodic SCL states. Although these models seem to provide a good fit
to $\tau _{T}^{(\exp )}$ (by using only one adjustable parameter, as the
fragility $m_{g}$ in the case of Ref.\onlinecite{Sch98}, or by using more
than three adjustable model parameters, as in the case of Ref.%
\onlinecite{KTZ96}), they need to be analyzed by considering the derivative
functions of the dynamical variables over the full temperature range.

As stressed by Angell et al. in Ref.\onlinecite{ANM00}, the popularity of
the thermodynamic AGM among molecular-liquid phenomenologists might be
compared with that of exploring the phenomenological VFT form by
glass-material experimentalists. The current study develops a statistical
approach for the description of CRRs, with their isobaric-isothermal
mechanism of solidification that were introduced a long time ago by Adam and
Gibbs in Ref.\cite{AG65}.

The application of the second-order derivative analysis for the case of
salol, revealed\cite{S95} that the widely used standard VFT form given in
Eq.(\ref{tau-1}) has no observation window. This observation ended the
unreserved domination of the phenomenological VFT form for fitting all
dynamical data of all SCLs, and divided the SCLs in two groups, ''regular''
and ''irregular'', that, respectively, obey or do not obey the VFT equation.
Meanwhile, it seems that all SCLs exhibit a jump in the timescale curvature
near $T_{c}$ (see e.g. Figs.\,7 and 8) related to the corresponding kink
observed in the temperature behavior of the timescale-steepness that
separates the glass forming liquids into two dynamically distinct states
(see e.g. Figs.\,5 and 6). Additionally, both the ''regular'' and
''irregular'' SCLs, therefore manifest the absolute thermodynamic
instability at the Kauzmann temperature $T_{K}$, whose physical meaning is
extended by the equation $T_{K}=T_{0}$.

We have discussed the SCLs in ergodic states in terms of isolated
finite-size solid-like clusters observed within the window $T_{g}\leq
T<T_{m} $ in the dynamic experiments through the primary relaxation
timescale data $\tau _{T}^{(\exp )}$. The cluster dynamics, introduced by
the Gaussian size-distribution cluster statistics and exponential-type
size-relaxation, described by Eqs.(\ref{Pz}) and (\ref{tauTz}),
respectively, is driven by the mesoscopic entropy and ensured by the Onsager
thermodynamic principle.

Below the crossover temperature $T_{c}$, the CRRs emerges as structurally 
{\em homogeneous clusters}, where thermodynamic stability is guaranteed by
the growth of the excess chemical potential $\Delta \mu _{g}$ ($>\Delta \mu
_{c}$). This energetic change, associated with the cluster rebuilding, is
justified experimentally through the analysis given in Table 2A. In turn,
the homogeneous clusters are associated with the strong SCL states of the
MCT, and/or with percolation theory solid-like clusters described by the VFT
equation (with the activation energy $E_{T}^{(SL)}=\Delta \mu
_{g}z_{T}^{(SL)}/T$ and the characteristic size $%
z_{T}^{(SL)}=z_{a}^{(SL)}/(1-T_{0}/T)$, both are given in Eq.(\ref{tau-SL}%
)). The minimum cluster size $z_{a}^{(SL)}$ is related to the {\em lattice
constant} of the underlying percolation lattice, i.e., $\zeta
_{a}^{(SL)}=\left( \text{v}_{0}z_{a}^{(SL)}\right) ^{1/3}$, where v$_{0}$ is
the molecule volume.

The experimental observation\cite{RA98} of the strong-SCL ergodic states,
described simultaneously by the thermodynamic AGM and the dynamic VFT
equations led to formulation of the thermodynamic-dynamic {\em %
glass-transition\ correspondence principle}. This principle is based on the
physically intelligible thermodynamic-dynamic equations introduced for the
critical $T_{0}$ and glass transition $T_{g}$ temperatures, as well as on
the ergodic hypothesis that gives rise to the temporal and spatial scale
minima introduced in Eq.(\ref{tau-and-z}). These scales are common for both
dynamic and thermodynamic experiments within the temperature domain $%
T_{g}\leq T<T_{c}$. Upon cooling near and below $T_{g}$, the cluster size $%
z_{T}^{(SL)} $ and its heterogeneous fluctuations $\Delta z_{T}^{(SL)}$ grow
and the SL clusters show an ergodic instability caused by their maximum
fluctuations\cite{K01}. The compact ''ergodic'' solid-like SL clusters are
therefore transformed into fractal-type clusters of the holey-like
(approximated by the ideal gas\cite{BS98,K01}) structure that is
accompaniment by expansion of the relevant percolation lattice, with $\zeta
_{b}>\zeta _{a}^{(SL)}$. Upon heating, the CRRs of the AGM disappear in the
dynamic experiments above $T_{c}$ but remain stable in calorimetric
thermally equilibrated experiments (see Fig.2). Moreover, as follows from
the first and second order $T$-derivative analyses elaborated in Refs.\cite
{S95,S96,RA98,thes}, the SL-cluster description given with the help of the
VFT-type forms is not useful for ''irregular'' SCLs such as salol, PDE, and
BMPC (see also steepness and curvature analyses in Figs.6A and 8, Fig.6B,
and Fig.6C, respectively).

Unlike the SL states, the {\em structurally heterogeneous clusters},
denominated as ML clusters, exhibit stability in all studied SCLs. They are
associated with excitations of the weak supercooled states, characteristic
of the moderately SCLs within the domain $T_{c}\leq T<T_{m}$ and introduced
by the VFTE equation with the correlation length $%
z_{T}^{(ML)}=z_{a}^{(ML)}/(1-T_{0}/T)$ described in Eq.\,(\ref{ET}). In
turn, the heterogeneous cluster-size fluctuations, $\Delta
z_{T}^{(ML)}=(2/\gamma _{c})^{1/2}$ $z_{T}^{(ML)}$, are introduced with the
help of the model parameter $\xi _{c}$ given in Eq.(\ref{Tc}). Due to the
common physical restriction imposed on fluctuations, which preserve the
ML-cluster structure and given by $\Delta z_{T}^{(ML)}/z_{T}^{(ML)}<1$, one
naturally arrives at the constraint for the timescale slowing-down exponent,
i.e., $\gamma _{c}>2$.

The minimum size of the ML clusters derived from the dynamical experimental
data (see Table 3) is related to the instability of heterogeneous clusters
below the crossover temperature $T_{c}$ that results in an expansion of the
underlying percolation lattice, i.e., $\zeta _{a}^{(SL)}>\zeta _{a}^{(ML)}$.
We see that within the context of percolation theory, the discussed cluster
reconstructions near $T_{g}$ and $T_{c}$ are geometrically similar and
therefore the crossover temperature $T_{c}$ for the moderately SCL states
plays a role similar to the ''crossover'' temperature $T_{g}$ for the strong
SCL states. This clarifies the effectiveness of the phenomenological HT-VFT
fitting form introduced on the basis of the VFT equation with relevant
substitution of the characteristic temperatures discussed in Eq.(\ref{Tc/T0}%
).

It has been justified experimentally (see analyses in Figs.\,5-8 ) that the
size fluctuations of the ML clusters, as ''non-rigid'' solid-like precursors
of the SL clusters, grow upon the decrease of the temperature. This means
that the ML clusters are characterized more as "fluid-like" rather than
"solid-like" and hence might be introduced on the basis of the thermodynamic
heterophase-fluctuation {\em microscopic} model proposed by Fischer and
Bakai in Ref.\onlinecite{FB99}. According to this model, the ML clusters may
be associated with{\em \ isolated liquid-like droplets }that show {\em small}
heterophase fluctuations. Upon cooling, the concentration of the solid
fraction grows and the liquid droplets exhibit a crossover from isolated to
non-isolated long-range density fluctuations\cite{B02}, related 
%interpercolating configurations, that are due 
to Fischer clusters that eventually do not contribute to the primary
relaxation. As seen from Fig.\,5 in Ref.\onlinecite{FB99}, for the case of
salol, the droplet percolation crossover temperature is close to $T_{c}$,
i.e. to that predicted by the MCT. The question therefore arises as to
whether the interpercolation droplet states, given by the Gibbs potential in
Eq.(2.1) in Ref.\onlinecite{FB99}, enable one to describe the drastic
temperature behavior observed for the timescale curvature\cite{S95} in the
window $T_{g}\leq T<T_{c}$ (see Fig.8). Furthermore, one can speculate that
the thermodynamic-dynamic glass-transformation correspondence above $T_{c}$,
can be established on the basis of the physical correspondence between the
thermodynamic heterophase-fluctuation droplet model\cite{FB99} and the
dynamic VFTE equation given for $\tau _{T}^{(ML)}$ in Eq.(\ref{tau-ML}).

One can expect that application of the derivative analysis by fitting the
timescale steepness and curvature, in addition to the common $\tau
_{T}^{(\exp )}$ fitting, might serve as a reliable instrument to give new
insights into important features of the underlying physical processes. This
is related to the weak--to-strong SCL-state crossover based on the existence
of the observation windows and the steepness kink discussed above in Eqs.(%
\ref{tau-and-z}) and (\ref{ineqs}), respectively. We exemplify our statement
by the Schulz model (SM)\cite{Sch98} that employs the concept of Goldstein's
energy landscape for SCLs. The overall ergodic-range timescale\cite{Sch98}
can be represented here by the activation energy $%
E_{T}^{(SM)}=E_{m}(T_{m}/T)^{\gamma _{g}}$ where $E_{m}$ and $\gamma _{g}$
are thermodynamic parameters (with $\gamma _{g}=1+2\Delta c^{(SM)}/3k_{B}$ ,
given in Eq.(26) in Ref.\onlinecite{Sch98} in terms of the excess
liquid-over-HT-solid isobaric specific heat capacity $\Delta c_{g}^{(SM)}$).
Ignoring the weakly $T$-dependent pre-exponential factor, one can estimate
the steepness function defined in Eq.(\ref{steep-func}) at $T_{g}$, i.e.,
the fragility $m_{g}$. Similar to the case of the VFT form, discussed on the
basis of Eq.(\ref{Tg/T0}), one can solve the system of the two equations for 
$m_{g}$ and $\tau _{g}$ with respect to $E_{m}$ and $\gamma _{g}$. This
process results in the following expression for the slowing-down exponent, $%
\gamma _{g}=$ $m_{g}/m_{g}^{*}$ , and offers a new equation for the Schulz
model excess heat capacity, namely 
\begin{equation}
\Delta c_{g}^{(SM)}=\frac{3}{2}(\frac{m_{g}}{m_{g}^{*}}-1)k_{B}\text{.}
\label{SM-predict}
\end{equation}
We have thereby demonstrated, along with the findings given in Eq.(\ref{Kok}%
) and analyzed in Fig.4 (for the self-consistent characteristic
temperatures) and in Eq.(\ref{delta-Sin}) (for the observed excess entropy $%
\Delta S_{\infty }$), how the steepness timescale analysis can be used to
verify the self-consistency of the model predictions. Both model Eqs.(\ref
{delta-Sin}) and (\ref{SM-predict}) need further testing.

We also speculate on the existence of some alternative mathematical forms
for the timescale $\tau _{T}$, whose cluster-size singular behavior near $%
T_{0}$ is sufficiently affected by the excess chemical potential $\Delta \mu
_{T}$. Indeed, in the current study, the macroscopic primary relaxation
timescale is introduced by the mesoscopic solid-like clusters with {\em \ }%
rearranging frequency $\omega _{T}^{(cl)}(z)\equiv \tau _{T}^{-1}(z)=\tau
_{a}\exp (-\Delta \mu _{T}$ $z/k_{B}T)$ (see Eq.(\ref{tauTz})). An
alternative description of the random mesoscopic rearranging regions was
proposed by Odagaki in Ref.\onlinecite{Oda95} on the basis of the AGM
treated within the controlled-diffusion approach. Formally, the rearranging
regions in SCLs were introduced as random domains defined by the {\em random}
excess chemical potential $\Delta \mu $ and by the characteristic size $%
z_{T}^{(AG)} $ that follows from Eq.(\ref{Sc-AG}) given by the AGM. The
domain rearranging frequency was therefore treated as a random value $\omega
_{T}^{(dom)}(\Delta \mu )=\tau _{d}^{-1}\exp (-\Delta \mu $ $%
z_{T}^{(AG)}/k_{B}T)$. In fact, the most general mathematical form is given
by $\omega _{T}\equiv \tau _{T}^{-1}(\Delta \mu ,z)=\tau _{a}^{-1}\exp
(-\Delta \mu $ $z/k_{B}T)$ where both $\Delta \mu$ and $z$ are random
variables. This type of extended generalized treatment might be considered
as a possible candidate to solve the problem of the primary relaxation
observed in the salol-type SCLs.

Finally, we have demonstrated how the primary relaxation scale can be
described by the two VFT and VFTE equations and parameterized by a finite
number of observables: three principal characteristic temperatures $T_{0}$, $%
T_{g}$ and $T_{c}$; the fragility $m_{g}$; the slowing-down exponent $\gamma
_{c}$; and the two strength indices $D^{(SL)}=\Delta \mu
_{g}z_{a}^{(SL)}/k_{B}T_{0}$ and $D^{(ML)}=\Delta \mu
_{c}z_{a}^{(ML)}/k_{B}T_{0}$ given in Eqs.(\ref{mT-SL}) and (\ref{mT-ML}),
respectively. Physically, this description is justified by the stabilization
of the strong and weak SCL states, realized through stabilization of the
temporal and spatial scales, and related to the minimum time and size of the
rearranging regions. This proposed approach is elaborated with the help of a
set of ''universal parameters'' $m_{g}^{*}=16\pm 2$ and $m_{c}^{*}=7\pm 1$,
that seems to be extended by the new ratio $z_{a}^{(SL)}/z_{a}^{(ML)}=4.2\pm
0.5$. Furthermore, one can show through application of the TD-D
correspondence principle, that CRRs are observed with the minimum
configurational entropy, i.e., $\Delta s_{g}^{(AG)}=s_{c}^{*}=k_{B}\ln (2\pm
0.25)$, as predicted by Adam and Gibbs in their pioneer work in Ref.%
\onlinecite
{AG65}. The new result will be discussed elsewhere.

ACKNOWLEDGMENTS

The authors are grateful to C. Austen Angell and Michael F. Shlesinger for
illuminating discussions. Thanks are due to Roberto Luis Moreira for
critical reading of the manuscript. Financial support by the CNPq (VKB)and
by the NSF-DMR-98-16257 (NSS) is also acknowledged.

\newpage

{\large \ FIGURE CAPTIONS}

.

Fig.\,1. First-order derivative function of the timescale, $10^{2}\left(
d\log _{10}(\tau _{T})/d(1/T)\right) ^{-1/2}$ versus temperature for
propylene carbonate. {\em Circles:} relevant dielectric relaxation and dc
conductivity experimental data $\tau _{T}^{(Exp)}$ from Fig.\,12 in Ref.%
\onlinecite{S96}. {\em Lines:} fitting forms proposed by the mode coupling
theory (MCT), Adam-Gibbs model (AGM) and phenomenological VFT ($T_{0}=132 K$%
, $B=389 K $) and the HT-VFT ($T_{0}^{(HT)}=153 K$, $B^{(HT)}=158 K$) forms.
One can see that the AGM and VFT forms fit the strongly SCL (SL) states,
shown between $T_{g}=157 K$ and $T_{c}=176 K$, and the HT-VFT form fits the
moderately SCL (ML) states and even normal liquid (NL) states above the
melting point $T_{m}=225 K$ . $T_{y}$ is the lower temperature of the MCT
form.

.

Fig.\,2. Typical temperature behavior of the configurational entropy of
supercooled liquids. Schematic representation of the thermodynamic data\cite
{RA98} (solid squares), reestimated dynamic dielectric relaxation data (open
circles, see Eq.(8) in Ref.\onlinecite{RA98}), and the mechanical stress
relaxation data (solid circles, see Fig.\,10 in Ref.\onlinecite{A91}) data.
The solid line corresponds to Eq.\,(\ref{Sc2}), which is common to the AGM
and VFT forms for strongly SCL states, and fits the thermally equilibrated
states within the whole ergodic region. Notations correspond to Fig.\,1.

.

Fig.\,3. Observation of the universal parameter $m_{g}^{*}$: scaled
characteristic temperatures against scaled fragility for glass forming
liquids (A) and polymers (B). Circles (A) and diamonds (B) correspond to
experimental data, see Tables 1A and 1B, respectively. Solid line obeys Eq.(%
\ref{Tg/T0}) with $m_{g}^{*}=16$.

.

Fig.\,4. Experimental justification of the characteristic-temperature
equations for supercooled liquids (A) and polymers (B). The symbols denote
experimental data from Table 1. Solid and broken lines are solutions of the
CTEs, (\ref{Kok}) with fixed ratio $m_{g}^{*}/m_{c}^{*}$, and (\ref{Oda}),
respectively. Dotted lines correspond to the upper and lower limits of Eq.(%
\ref{Kok}) subject to the condition $m_{g}^{*}/m_{c}^{*}=2$.

.

Fig.\,5. Steepness-temperature analysis of the primary timescale in
supercooled liquids propylene carbonate (A), ortho-terphenyl (B) and
1-propanol (C) where the VFT equation (\ref{tau-VFT-equation}) is
experimentally justified. Symbols correspond to the dynamical data
(dielectric relaxation, dc conductivity, and viscosity) by Stickel et al.
reestimated with the help of Eq.(\ref{F1-T}) from Fig.12 in Ref.%
\onlinecite{S96}, from Fig.\,6 in Ref.\onlinecite{S97} and from Fig.15 in
Ref.\onlinecite{S96} for the cases A, B and C, respectively. Lines are given
by the VFT and VFTE equations using the experimental characteristic
temperatures from Table 1 and with the adjustable parameters given in Eqs.(%
\ref{mT-SL}) and (\ref{mT-ML}), respectively. Notations correspond to
Fig.\,1.

.

Fig.\,6. Steepness-temperature analysis of the primary timescale in
supercooled liquids salol (A), phenolphthalein-dimethyl-ether (B) and
bis-methoxy-phenyl-cyclohexane (C) where the applicability of the VFT
equation (\ref{tau-VFT-equation}) is questioned. Symbols correspond to the
dynamical data given by Stickel et al. and reestimated with the help of Eq.(%
\ref{F1-T}) from Fig.\,4 in Ref.\onlinecite{S95}, from Fig.\,10 in Ref.%
\onlinecite{S96} and from Fig.\,7 in Ref.\onlinecite{S97} for the cases A, B
and C, respectively. Lines and fitting parameters correspond to Fig.\,5. The
VFT fit is questionable at low temperatures.

.

Fig.\,7. Curvature-temperature analysis for the primary relaxation in
propylene carbonate. Symbols correspond to experimental data reestimated
using Eqs.(\ref{curva}) and (\ref{F2-VFTE}) with the help of Fig.13 in Ref.%
\cite{S96}. Lines are given by the VFT equation ($k_{T}^{(VFT)}$ ) for the
strongly SCL state (SL), HT-VFTE ($k_{T}^{(HT-VFT)}$ ) and VFTE ($%
k_{T}^{(VFTE)}$ ) equations for the moderately SCL state (ML), all discussed
in the text, with the same adjustable parameters as found in Fig.\,5A.

.

Fig.\,8. Curvature-temperature analysis for the primary relaxation in salol.
Symbols correspond to experimental data reestimated from Fig.8 in Ref.%
\onlinecite
{S95}. Lines represent equations given in the text with the same adjustable
parameters as found in Fig.\,6A.

\newpage 


\begin{references}
\bibitem{ANM00}  C.A. Angell, K.L. Ngai, G.B. McKenna, P.F. McMillan, and
S.W. Martin, J. Appl. Phys. {\bf 88}, 3113 (2000).

\bibitem{BNA93}  R. B\"{o}hmer, K.L. Ngai, C.A. Angell, and D.J. Plazek, J.
Chem. Phys. {\bf 99}, 4201(1993).

\bibitem{A91}  C.A. Angell, J. Non-Cryst. Sol. {\bf 131-133}, 13 (1991).

\bibitem{GG92}  W. G\"{o}tze, L. Sj\"{o}gen, Rep. Prog. Phys. {\bf 55}, 289
(1992).

\bibitem{Bin99}  K. Binder {\it et al}., in {\it Slow Dynamics in Complex
Systems: Eighth Tohwa University International Symposium }(M. Tokuyama, I.
Oppenheim, AIP, 1999), p.193.

\bibitem{KTB98}  A. Kudlik, C. Tschirwitz, T. Blochowicz, S. Benkhof, and E.
R\"{o}ssler, J. Non-Cryst. Solids {\bf 235-237}, 401 (1998).

\bibitem{S95}  F. Stickel, E.W. Fischer and R. Richert, J. Chem. Phys. {\bf %
102}, 6251 (1995).

\bibitem{S96}  F. Stickel, E.W. Fischer, and R. Richert, J. Chem. Phys. {\bf %
104}, 2043 (1996).

\bibitem{S97}  C. Hansen, F. Stickel, T. Berger, R. Richert, and E.W.
Ficher, \ J. Chem. Phys. {\bf 107}, 1086 (1997).

\bibitem{RA98}  R. Richert and C.A. Angell, J. Chem. Phys. {\bf 108}, 9016
(1998).

\bibitem{KTZ96}  D. Kivelson, G. Tarjus, X. Zhao, and S. A. Kivelson, Phys.
Rev. E{\bf \ 53}, 751 (1996).

\bibitem{Sch98}  M. Schulz, Phys. Rev. B57, 11319 (1998).

\bibitem{KKZ95}  D. Kivelson, S. A. Kivelson, X. Zhao, Z. Nussinov, and G.
Tarjus, Physica A{\bf \ 219}, 27 (1995).

\bibitem{AG65}  J. H. Gibbs and G. Adam, J. Chem. Phys. {\bf 43},139 (1965).

\bibitem{CG79}  M.N. Cohen and G.S. Grest, Phys.Rev. B{\bf \ 20}, 1077
(1979).

\bibitem{WS87}  T.A. Weber and F.H. Stillinger, Phys. Rev. B{\bf \ 36}, 7043
(1987)

\bibitem{SSH91}  J.P. Sethna, J.D. Shore, and M. Huang, Phys. Rev. B{\bf \ 44%
} 4943, (1991).

\bibitem{NRP91}  K.L. Ngai, R.W. Rendell, and D.J. Plazek, J. Chem. Phys. 
{\bf 94}, 3018 (1991).

\bibitem{Oda95}  T. Odagaki, Phys. Rev. Let. {\bf 75}, 37 (1995).

\bibitem{Bak98}  A.S. Bakai, Low Temp.\ Phys. {\bf 24}, 20 (1998).

\bibitem{Cha99}  R.V. Chamberlin, Phys. Rev. B{\bf \ 48} 15638 (1993); Phys.
Rev. Lett. {\bf 82}, 2520 (1999).

\bibitem{FB99}  E.W. Fischer and A.S. Bakai, in {\it Slow Dynamics in
Complex Systems: Eighth Tohwa University International Symposium }(M.
Tokuyama, I. Oppenheim, AIP, 1999), p. 325.

\bibitem{K99}  V.B. Kokshenev, Physica A{\bf \ 262}, 88 (1999).

\bibitem{A97}  C.A. Angell, Physica D{\bf \ 107}, 122 (1997); J. Res. Natl.
Inst. Stand. Technol. {\bf 102},171 (1997).

\bibitem{NRM96}  V.N. Novikov, E. R\"{o}ssler, V.K. Malinovsky, and N.V.
Surovtsev, Europhys. Lett. {\bf 35}, 289 (1996); E.R\"{o}ssler, A.P.Sokolov,
Chem. Geol. {\bf 128}, 143 (1996).

\bibitem{Isi92}  M.B. Isichenko, Rev. Mod. Phys. {\bf 64}, 961 (1992).

\bibitem{BS98}  J.T. Bendler and M.F. Shlesinger, J. Stat. Phys. {\bf 53},
531 (1998).

\bibitem{BFS01}  J.T. Bendler, J.J. Fontanella and M.F. Shlesinger, Phys.
Rev. Lett. {\bf 87}, N19 (2001).

\bibitem{K01}  V.B. Kokshenev, Solid State Commun. {\bf 119}, 429 (2001).

\bibitem{Cha96}  R.V. Chamberlin, Europhys Lett.{\it \ }{\bf 33}, 545 (1996).

\bibitem{CHA93}  R.V. Chamberlin, Phys. Rev. B{\bf \ 48}, 15 638 (1993).

\bibitem{K98}  V.B. Kokshenev, Phys. Rev. E{\bf \ 57}, 1187 (1998).

\bibitem{KS01J}  V.B. Kokshenev and N.S. Sullivan, J. Low Temp. Phys. {\bf %
122}, 221 (2001).

\bibitem{KS01}  V.B. Kokshenev and N.S. Sullivan, Phys. Lett. A{\bf \ 208},
97 (2001).

\bibitem{CBS92}  R.V. Chamberlin, R. B\"{o}hmer, E. Sanchez, and C. A.
Angell, Phys. Rev. {\bf B46}, 5787 (1992).

\bibitem{NM67}  A. Neapolitano and P.B. Macedo, J. Res. Natl. Bur. Stand. 
{\bf A71} , 231 (1967);

\bibitem{Pri80}  V. Privalko, J. Chem. Phys. {\bf 84}, 3307 (1980).

\bibitem{AS76}  C.A. Angell and W. Sichina, Ann. N.Y. Acad. Sci., {\bf %
279-280}, 53 (1976).

\bibitem{A95}  C.A. Angell, Science {\bf 267}, 1924 (1995).

\bibitem{Sou90}  J. Souletie{\it ,} J. Phys. France {\bf 51}, 883 (1990).

\bibitem{Nga94}  K.L Ngai, in: R. Richert, A. Blumen (Eds.), Disorder
Effects on Relaxation Processes: Glasses, Polymers, Proteins, Springer, p.89
(1994).

\bibitem{Bra94}  A.J. Bray, Adv. Phys. {\bf 43}, 57 (1994).

\bibitem{fragile}  Many authors use this equation in the inversely form $%
m_{g}=m_{g}^{*}(1+\varepsilon _{g}^{-1})$ with $\varepsilon
_{g}=T_{g}/T_{0}-1$ to evaluate the fragility by the known characteristic
temperatures $T_{g}$ and $T_{0}$.

\bibitem{precise}  More precise calculations are given in Table I in Ref.%
\cite{RA98}, $m_{1}^{*}=2-A$, that results in $m_{1}^{*}=16\pm 3$.

\bibitem{BA92}  R. B\"{o}hmer and C.A. Angell, Phys. Rev. B{\bf \ 45}, 1009
(1992).

\bibitem{define}  One of the authors uses the opportunity to correct here a
denomination of $m_{T}$ as the steepness function and not as the ''fragility
function'' employed in Ref.\onlinecite{K99}.

\bibitem{cluster}  Formally, $z$ is a random-site occupation number of the
underlying percolative lattice.

\bibitem{L-L}  L.D. Landau and E.M. Lifshitz, Statistical Physics, Pergamon
Press, London (1989).

\bibitem{onsager}  For Onsager's principle see e.g. Eq.(112.1) in Ref.%
\onlinecite
{L-L}.

\bibitem{model}  An alternative expression to that given in Eq.(\ref{CRR-T})
was proposed in Eq.(2.3) in Ref.\onlinecite{FB99} for the reduced $z_{T}$
that fits well calorimetric experimental data for salol.

\bibitem{TYM95}  S. Takahara, O. Yamamuro, T. Matsuo, J. Phys. Chem. {\bf 99}
9589 (1995).

\bibitem{YIT99}  O. Yamamuro, M. Ishikawa, I. Tsukushi, and T. Matsuo, in
Slow Dynamics in Complex Systems: Eighth Tohwa University International
Symposium, edited by Michio Tokuama and Irwin Oppenheim, AIP, p.513 (1992).

\bibitem{GR}  I.S. Grandshteyn and I.M. Ryzhik, {\it Table of Integrals,
Series, and Products}, Ed. Alan Jeffrey, (1994).

\bibitem{trial}  Within this context the trial estimations of the steepness
function given through Eq.(\ref{m-SCL}) are considered as rough estimations
for liquids, which ignore this kink. For the case of polymers the existence
of the kink is not clear.

\bibitem{thes}  P.D. Borges, {\it Cluster-Correlation Approach to the
Primary Relaxation Near the Glass Transition}, M.S. Thesis, Universidade
Federal of Minas Gerais, 2001, unpublished.

\bibitem{B02}  A.S. Bakai, J. Non-Cryst. Solids, in press (2002).
\end{references}
\end{document}